\newcommand{\be}{\begin{equation}}
\newcommand{\ee}{\end{equation}}

\documentclass[aps,prl,twocolumn]{revtex4}
\usepackage{graphicx}
\usepackage{amssymb}
\usepackage{bm}

\begin{document}

\title{Localization of Two-Component Bose-Einstein Condensates in Optical Lattices}

\author{Elena A. Ostrovskaya and Yuri S. Kivshar}

\affiliation{Nonlinear Physics Group, ARC Centre for
Quantum-Atom Optics and Research School of Physical Sciences and
Engineering, Australian National University, Canberra ACT 0200,
Australia}

\date{\today}
\begin{abstract}
We reveal underlying principles of nonlinear localization of a
two-component Bose-Einstein condensate loaded into a
one-dimensional optical lattice. Our theory shows that
spin-dependent optical lattices can be used to manipulate both the
type and magnitude of nonlinear interaction between the ultracold
atomic species and to observe nontrivial two-componentnlocalized states of a
condensate in both bands and gaps of the matter-wave band-gap structure.
\end{abstract}
\maketitle

Recent experimental studies of Bose-Einstein condensates (BECs) in
optical lattices, including Bloch-band
spectroscopy~\cite{phillips_bands}, patterned loading of the
lattice~\cite{phillips_loading}, and
nonlinear tunnelling~\cite{arimondo_tunneling},
demonstrate unprecedented levels of manipulation of
the coherent matter waves. Tunability of the periodic lattice
potential combined with intrinsic nonlinearity of BEC provides an excellent testbed for many concepts of
condensed matter physics and nonlinear physics. In particular, strong parallels have been drawn \cite{our_prl} between nonlinear optics
of ultracold atoms in optical lattices and behaviour of optical waves in periodic nonlinear media, where instability and localization
phenomena have been well explored \cite{optics_nature}.  Akin to optical structures with periodic refractive index, such as waveguide arrays and photonic crystals, optical lattices form a band-gap structure for coherent matter waves
which modify diffraction properties of atomic wavepackets \cite{effective_mass}. The
coexistence of both normal and anomalous diffraction regimes was
predicted to lead to a nonlinear localization of condensate with
both attractive and repulsive interactions in the gaps of a linear
Bloch-wave spectrum~\cite{lenaPRA}. Recent experiments on
diffraction management of BEC in lattices
\cite{marcus_dm,inguscio_dm} have paved the way to the first
observation of a localized atomic wavepacket (bright gap soliton) in a
BEC with repulsive interactions~\cite{gap_observation}.

Most of the theoretical and experimental studies of coherent
matter waves in optical lattices are concerned with
single-component BECs. However, it is well known that
condensate mixtures display many novel physical effects not
found in single-species BECs~\cite{two_species}, including domain
separation of the ground state and stabilization of nonlinear
localized states. Experiments on spin-dependent
optical lattices~\cite{spin_dependent} have shown that a
two-component BEC composed of two hyperfine states of the same
atomic species can be effectively and coherently manipulated in an
optical lattice. Recently, analysis of periodic (Bloch) states of a multi-component BEC in a lattice \cite{two_deconinck,two_salerno} has revealed that the modulational instability of excited periodic states can potentially lead to the formation of multi-component localized states (solitons) \cite{two_salerno}. This conclusion agrees with the recent theoretical study of nonlinear inter-gap localization of multi-component coherent light fields in periodic
optical structures \cite{andrey_mgap}. 

In this Letter, we address the problem of
nonlinear localization of a two-component condensate and formation
of two-component matter-wave solitons in an optical lattice.
We show that an optical lattice can be used to modify the type of
the effective nonlinear interactions both within and between the
condensate species, {\em without the Feshbach resonance
manipulation} of the scattering lengths. As a consequence, novel
types of nonlinear localization of coherent matter waves can be
achieved both in gaps and bands of the linear Bloch-wave spectrum.
In particular, when one of the condensate components is in the
periodic Bloch state, the second component exhibits effective
periodic potential combined with the optical potential and that
induced by the mean-field of the Bloch wave. As a result,
nontrivial localization of a two-component BEC in a {\em
self-induced lattice} can occur in the form of a two-component
``dark-bright'' band-gap soliton. The effects of two-component
localization discussed here can be explored in the current BEC
experiments, as well as in the optically-induced photonic
lattices~\cite{optics_nature}.

\begin{figure}
\centerline{\scalebox{0.46}{\includegraphics{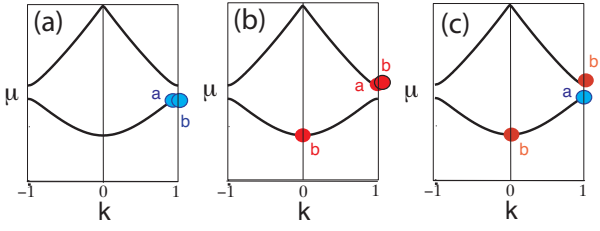}}}
\caption{Chemical
potential and momentum of the BEC components, relative to the lattice band-gap
structure, necessary to achieve
(a) attractive-attractive, (b) repulsive-repulsive, and (c)
repulsive-attractive effective nonlinear interactions. For the fixed parameters of the $|a\rangle$ component either of the $|b\rangle$ locations can be chosen. Colors code different signs of effective diffraction.} \label{fig1}
\end{figure}

We model the dynamics of a two-component BEC with {\em repulsive} interactions in an optical lattice
by the mean-field equations for the wavefunctions of the
condensate components $|a\rangle$ and $|b\rangle$:
\begin{equation}\label{model}
i\frac{\partial}{\partial t}\begin{pmatrix}{\Psi_a \cr \Psi_b}\end{pmatrix} =\begin{pmatrix}
  {\hat{L_a}    &  \hat{L}_{ab} \cr
   \hat{L}^*_{ab}    &   \hat{L}_b}
\end{pmatrix} \begin{pmatrix}{\Psi_a \cr \Psi_b}\end{pmatrix},
\end{equation}
where $\hat{L}_{n}=-\partial^2/\partial
x^2+V_{n}(x)+g_{nn}|\Psi_{n}|^2$, and
$\hat{L}_{ab}=g_{ab}\Psi_a\Psi^*_b$. Here we assume that the two
BEC components can exhibit different potentials, namely:
$V_a(x)=V_0\sin^2(k_Lx)$ and $V_b(x)=V_0\sin^2(k_Lx+\theta)$, which can be
shifted relative to each other, e.g. by varying the polarization
angle in the case of a spin-dependent lattice \cite{spin_dependent}. These potentials are less constrained than those in \cite{spin_dependent} to allow for consideration of a general setup where the condensate components may actually belong to different atomic species \cite{Rb_K}, being manipulated independently. The one-dimensional model (\ref{model}) is derived for strongly anisotropic BEC clouds with the tight confinement
direction transverse to that of the lattice~\cite{lenaPRA} and
made dimensionless by adopting the lattice units of length
$x_0=k^{-1}_L$ (typically $\sim 10^{-1}~\mu {\rm m}$) and energy $E_r=\hbar^2k^2_L/2m$. The stationary
states of the wavefunction of the $n$-th BEC component are found as
$\Psi_n(x,t)=\psi_n(x) \exp(-i\mu_n t)$, where $\mu_n$ is the
chemical potential normalized by $E_r$. Without the repulsive mean-field interactions, the ground and excited states of each of the BEC components
are periodic matter waves: $\psi_n=\phi_n(x)\exp(ikx)$, where the
lattice momentum $k$ lies within the Brillouin zone. The
periodicity of the lattice leads to the band-gap structure of the
spectrum $\mu_n(k)$. At the $j$-th edge of the Brillouin zone $\mu_n\equiv \mu^{\{j\}}_n$, the wavefunction 
 $\phi^{\{j\}}_n(x)=\phi^{\{j\}}_n(x+\pi/k_L)$ is a Bloch state with the lattice periodicity. Schematics of the
Bloch-wave spectrum in the reduced-zone representation ($|k|<1$)
is shown in Fig.~\ref{fig1}.

A stationary BEC wavepacket located, in the momentum $k-$space,
near the $j$-th band edge can be described as a corresponding Bloch wave, dressed by a slowly varying (in $x$) envelope,
$\Psi_n(x,t;\mu^{\{j\}}_n)=\Phi_n(x,t)\phi^{\{j\}}_n(x)$. Considering that the
condensate components $|a\rangle$ and $|b\rangle$ can be
positioned at the band edges $i$ and $j$, respectively, the
envelope equations take the form~\cite{andrey_mgap}:
\begin{eqnarray}
\label{envelopes}
-i\frac{\partial \Phi_a}{\partial t}=\left[\frac{1}{2}D^{\{i\}}_a
\frac{\partial^2}{\partial
x^2}-{\tilde{g}}^{\{ii\}}_{aa}|\Phi_{a}|^2-\tilde{g}^{\{ij\}}_{ab}|\Phi_{b}|^2\right]
\Phi_a\cr -i\frac{\partial \Phi_b}{\partial
t}=\left[\frac{1}{2}D^{\{j\}}_b \frac{\partial^2}{\partial
x^2}-{\tilde{g}}^{\{jj\}}_{bb}|\Phi_{b}|^2-\tilde{g}^{\{ij\}}_{ba}|\Phi_{a}|^2\right]
\Phi_b,
\end{eqnarray}
where the effective diffraction of the wavepacket is defined by
the curvature of the corresponding band edge,
$D^{\{i,j\}}_{a,b}=
\left(\partial^2 \mu_{a,b}/\partial
k^2 \right )_{\mu_{a,b}=\mu^{\{i,j\}}_{a,b}}$, and the effective interaction coefficients depend on the
symmetries of the Bloch states at the band edges:
$\tilde{g}^{\{ij\}}_{ab}={g}^{\{ij\}}_{ab}\int|\phi^{\{j\}}_a\phi^{\{j\}}_b|^2dx$.
The relative shift of the spin-dependent lattice
potentials can enhance or dampen the cross-species nonlinear
interaction without any change in the effective diffraction, which
introduces a new degree of experimental control into the system. Eqs. (\ref{envelopes}) and their
complex conjugate form are equivalent, so even for the case of anomalous diffraction ($D_n<0$)~\cite{effective_mass} the equations can always be
re-written in the more intuitive normal diffraction form, but with
opposite signs of the effective nonlinear coefficients.

Envelope equations (\ref{envelopes}) predict {\em three types of
the effective nonlinear interactions} that can be realized for
two-component BECs in a lattice. These regimes can be achieved by
appropriate placing of the BEC wavepackets parameters relative to the band structure, as
illustrated schematically in Figs.~\ref{fig1}(a-c). Below we
consider only the main band ($B1$: $\mu^{\{1\}}|_{k=0}\leq \mu \leq \mu^{\{2\}}|_{k=1}$) and the first excited band ($B2$:
$\mu^{\{3\}}|_{k=1} \leq \mu \leq \mu=\mu^{\{4\}}|_{k=0}$) of the lattice that can be populated in
experiments~\cite{phillips_bands}.

\begin{figure}
\centerline{\scalebox{0.8}{\includegraphics{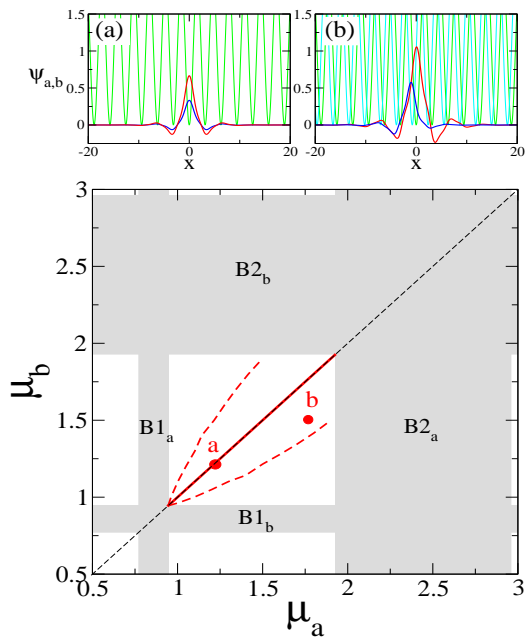}}}
\caption{Existence domain for the bright-bright
atomic gap solitons in the $\{\mu_a,\mu_b\}$ plane ($V_0=2$, $g_{aa}=g_{bb}=g_{ab}=1$). Shaded
-- two lowest Bloch bands $B1_{a,b}$ and $B2_{a,b}$
exhibited by the condensate components $|a\rangle$ and
$|b\rangle$. Solid line - existence domain $\mu_a=\mu_b$ predicted
by the envelope theory and linear waveguiding calculations (see
text). Dashed lines - borders of the existence domain for the
relative lattice shift $\theta=\pi/4$. Top panels: examples
of the condensate wavefunctions at the marked points of the existence
domain.} \label{fig2}
\end{figure}

{\em Attractive-attractive} interaction regime [Fig.~\ref{fig1}(a)] can be realized when
$\tilde{g}_{aa,bb},\tilde{g}_{ab,ba}<0$,
$|\tilde{g}_{aa}|=|\tilde{g}_{bb}|=|\tilde{g}_{ab}|=|\tilde{g}_{ba}|$.
In this regime, near the band edge $\mu=\mu^{\{2\}}|_{k=1}$ both
condensate components exhibit anomalous diffraction, and therefore
can exhibit self-focusing in the form of {\em two-component bright gap
solitons}. The envelope equations (\ref{envelopes}) take the form
of the integrable Manakov system, well studied in the context of
nonlinear optics (see, e.g., \cite{dark_bright_optics}). According to Eq. (\ref{envelopes}), both localized components can be treated as ground states of the same effective potential; they have equal
chemical potentials $\mu_a=\mu_b$ and widths of the spatial
density distribution envelopes: $\Phi_a(x)=\Phi_0(x)\cos\alpha$,
$\Phi_b(x)=\Phi_0(x)\sin\alpha$, where
$\Phi_0(x)=\sqrt{\mu_a/\tilde{g}_{aa}}{\rm sech}(\sqrt{2\mu_a}x)$,
and $\alpha$ is an arbitrary parameter.  The existence
domain for two-component solutions in the parameter space $\{\mu_a,\mu_b\}$ can also be calculated using linear waveguiding principles. Assuming that the
densities of the two component are largely dissimilar, one can
decouple Eqs. (\ref{model}) and find, numerically, spatial
profiles of single-component bright gap solitons $\psi_a(x;\mu_a)$
in the entire gap $\mu^{\{2\}}_a\leq \mu_a \leq \mu^{\{3\}}_a$.
The low-density second component, $\psi_b$ can then be found as a ground
state (i.e. lowest-order guided mode) of the {\em effective linear
potential} combined of the lattice potential and the
mean-field potential induced by the $|a\rangle$-component. The
cut-off values $\mu_b(\mu_a)$ for this ground state define the
lower boundary of the existence domain for two-component localized
states of the condensate, and the upper boundary is $\mu_b=\mu_a$. The
linear waveguiding analysis of the complete model (\ref{model})
confirms the prediction of the envelope theory that in the
degenerate case of equal effective interaction coefficients the
existence domain is confined to the line $\mu_b=\mu_a$ (see Fig.
2). Remarkably, this degeneracy can be lifted by shifting
the lattices trapping the two condensate components relative to each other. The shift destroys the
equality between the interaction coefficients so that
$\tilde{g_{ab}}/\tilde{g}_{aa}>1$ and, according to the envelope
theory, yields the new cut-off values for the $\psi_b$ modes,
$\mu_b=(\mu_a/4)(1-\sqrt{1+8\tilde{g}_{ab}/\tilde{g}_{aa}})^2 >
\mu_a$. The boundaries of the significantly expanded existence
domain calculated from the guided-mode analysis of the complete
model (\ref{model}) are shown in Fig.~\ref{fig2} by dashed lines;
such a dramatic expansion can be achieved by the relatively small
shift of $\theta=\pi/4$.

\begin{figure}
\centerline{\scalebox{0.8}{\includegraphics{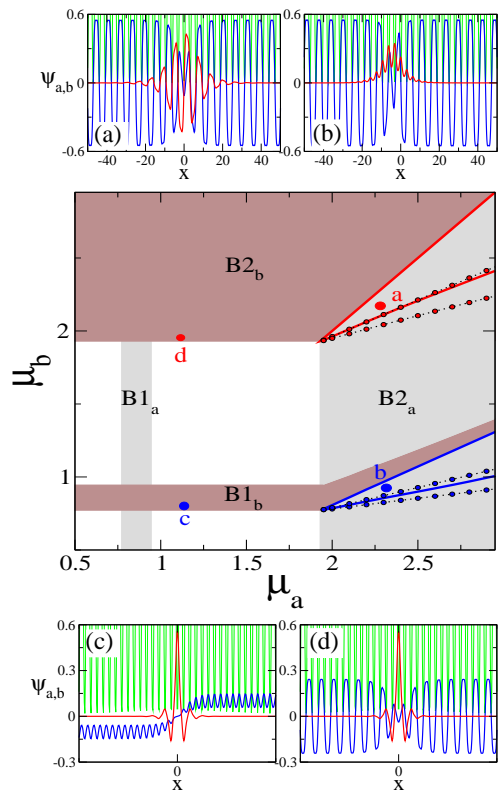}}}
\caption{Existence domain for the dark-bright
atomic solitons ($V_0=2$, $g_{aa}=g_{bb}=g_{ab}=1$). Shaded -- two lowest Bloch
bands $B1_{a,b}$ and $B2_{a,b}$ exhibited by the condensate
components $|a\rangle$ and $|b\rangle$. Solid lines - borders of
the existence domains in repulsive-repulsive regime derived from the linear waveguiding
calculations (see text). Dotted lines - 
borders of the existence domains for the
lattices shift $\theta=\pi/4$. Top (bottom) panels: examples of the
condensate wavefunctions in the repulsive-repulsive (attractive-repulsive) regime at the marked points of the existence
domains.} \label{fig3}
\end{figure}

\begin{figure}
\centerline{\scalebox{0.52}{\includegraphics{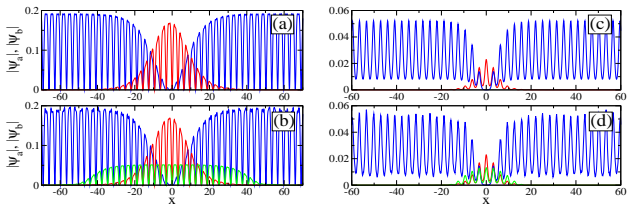}}}
\caption{Temporal evolution of a dark-bright
localized state in the repulsive-repulsive regime with both components within (a,b) the second band, at $\mu_a=2.2$,
$\mu_b=2.1$, and (c,d) the first band at $\mu_a=0.81$, $\mu_b=0.8$.
Shown are (a,c) initial spatial density profiles and profiles of
the coupled state and a spreading bright component decoupled from the dark
one at (b) $t=1.35$ mc and (d) $t=2.7$ mc. Note the slower
expansion for the lower density state (c).} \label{fig4}
\end{figure}

{\em Repulsive-repulsive} regime [Fig.~\ref{fig1}(b)] can be
achieved when $\tilde{g}_{aa,bb},\tilde{g}_{ab,ba}>0$,
$|\tilde{g}_{ab}|=|\tilde{g}_{ba}|$, and the wavepackets are
located either at the same, $|\tilde{g}_{aa}|=|\tilde{g}_{bb}|$,
or different, $|\tilde{g}_{aa}|\neq |\tilde{g}_{bb}|$, band edges.
Let us first consider the BEC wavepacket located at the edge of the second
band ($B2_a$), $\mu_a=\mu^{\{3\}}_a|_{k=1}$. It experiences normal diffraction
and can support a {\em dark soliton} with the envelope
$\Phi_a=\sqrt{\mu_a/\tilde{g}_{aa}}{\rm tanh}(\sqrt{\mu_{aa}} x)$,
imprinted onto the extended background of a Bloch wave 
$\phi^{\{3\}}_a(x)$ ~\cite{in_band}. Originating at the
band edge, such dark states with zero group velocity can exist
within the entire band. Their spatial structure
suggests that they can be created experimentally by
phase-imprinting technique, which does not require access to the
gap. It is easy to see that, in a two-component BEC, the Bloch
state in $|a\rangle$-component induces a periodic potential for
the $|b\rangle$ component which acts together with the potential
of the optical lattice. As a result, the band-gap structure of the
Bloch-wave spectrum for the $|b\rangle$-component is significantly
modified. The original bands are shifted, so that the edge of the modified second band $\mu^{\{3\}}_b$ coincides
with the chemical potential of the {\em nonlinear} Bloch state $\phi^{\{3\}}_a$ (see Fig.~\ref{fig3}). The remarkable effect of this condensate-induced
lattice potential is that the $|b\rangle$-component can now be
spatially localized in every gap of the {\em induced band-gap
structure} in the form of a {\em bright gap soliton}. Moreover, the
existence domains for the coupled states calculated using the
waveguiding properties of the dark in-band solitons, lie entirely within the original bands
$B1$ and $B2$. This type of
localization could have a dramatic experimental manifestation,
whereby formation of a bright soliton in one of the condensate
components can be achieved by {\em phase-imprinting onto the Bloch
state of the complementary component} in the spectral band. This effect can be observed when both components are
located either at the same or the opposite band edges [Fig. 1(b)], but always in the normal diffraction regime. The localization of a single-component repulsive BEC in the form of a bright soliton
is impossible in this regime. Therefore, although the 
dark-bright state can be dynamically stable [see
Figs.~\ref{fig4}] for a dark state imposed onto both modulationally stable [Fig.\ref{fig4} (c)] and unstable [Fig.\ref{fig4} (a)] Bloch-wave backgrounds, decoupling of the condensate components leads to rapid spread of the bright localized
state, proportional to its mean-field energy [Figs.~\ref{fig4}(b,d)].

Dark-bright solitons supported by effectively repulsive nonlinear
interactions have been previously
discussed in the context of nonlinear
optics~\cite{dark_bright_optics} and BEC in a harmonic
potential~\cite{dark_bright_bec}. The striking difference between
the dark-bright localized states in the optical lattice and
dark-bright BEC solitons in a harmonic potential is that the bright
component in the lattice is localized in the gaps of the
{\em induced} band-gap structure. Therefore the multiple existence
domains for the two-component state always lie between the cut-off line of the ground mode
guided by a dark state and the corresponding {\em edge of the
induced band}. In contrast with the case of bright-bright gap solitons,
these existence domains shrink (dotted lines in Fig. 3) with the relative shift between the lattices $V_a(x)$ and $V_b(x)$,
due to reduced effect of the induced periodic potential on the bands shift. Another drastic
feature of the dark-bright states in the lattice is the energy difference
between the states centered at the lattice minima and maxima,
which inhibits mobility of the localized states
across the lattice. As a result, the interaction of two
dark-bright states with zero relative velocity
(which, without the lattice, depends of the relative phase of the bright components \cite{dark_bright_optics}) is suppressed, and their spatial separation can be constant with time.

{\em Repulsive-attractive} regime [Fig.~\ref{fig1}(c)] can be
accessed when $\tilde{g}_{aa},\tilde{g}_{ab}<0$,
$\tilde{g}_{bb},\tilde{g}_{ba}>0$,
$|\tilde{g}_{ab}|=|\tilde{g}_{ba}|$, and $|\tilde{g}_{aa}|\neq
|\tilde{g}_{bb}|$. The BEC components should be located at different band edges, which is more difficult to arrange experimentally. However, in the absence of optical lattice, this regime could only be explored for a mixture of BECs with opposite signs of the scattering lengths. In this
regime, the envelope equations (\ref{envelopes}) predict the
existence of ``normal'' and ``reverse'' pairs of dark-bright
solitons~\cite{symbiotic}. In the ``normal'' case, a dark soliton on
the Bloch-wave background in the normal diffraction regime (e.g.,
at $\mu_a\geq\mu^{\{3\}}_a$) is coupled to a bright gap state in the
anomalous diffraction regime (e.g., at $\mu_b\geq\mu^\{2\}_b$). Both
components can exist independently as dynamically stable localized
states, and the coupling is achieved for
$\tilde{g}^2_{ab}<|\tilde{g}_{aa}\tilde{g}_{bb}|$. Typical examples of condensate wavefunctions in this regime are shown in Fig. 3(c,d). The bright localized component $|a\rangle$ exists in the entire gap, whereas the dark $|b\rangle$ component forms a coupled state only in the vicinity of the band edge.  Although  in the homogeneous case the two-component state is highly unstable with respect
to the mutual displacement of its constituents~\cite{symbiotic}, the optical lattice potential has a stabilizing effect on its dynamics. The ``reverse'' combination,
when the dark state in the anomalous diffraction regime
(effectively attractive condensate) is coupled to a bright state
in the normal diffraction regime (effectively repulsive
condensate) requires that
$\tilde{g}^2_{ab}>4|\tilde{g}_{aa}\tilde{g}_{bb}|$~\cite{symbiotic},
which is hard to achieve for realistic experimental conditions. 


In conclusion, we have analyzed the 
two-component BEC in optical lattices and shown that
spin-dependent lattices can be used to vary both the type and
magnitude of nonlinear interactions without manipulating atomic
scattering lengths. Using a simple envelope theory combined with the
principles of linear waveguiding, we have demonstrated that
localization of BEC in the form of gap solitons can occur both in
the gaps and bands of the linear Bloch-wave spectrum. The formation of
the dark-bright composite state is a striking effect that does not require
accessing the spectral gaps, which makes it attractive for possible experimental observations.

 We thank M. Oberthaler for many fruitful discussions.

\end{document}